\documentclass[graybox]{svmult}

\usepackage{mathptmx}       
\usepackage{helvet}         
\usepackage{courier}        
\usepackage{type1cm}        
\usepackage{makeidx}         
\usepackage{graphicx}        
\usepackage{multicol}        
\usepackage[bottom]{footmisc}
\usepackage{hyperref}

\makeindex             

\begin{document}

\title*{The HOG-FDA Approach with Mobile Phone Data to Modeling the Dynamic of People's Presences in the City}
\titlerunning{The HOG-FDA Approach with Mobile Phone Data} 
\subtitle{\emph{L'approccio HOG-FDA con i dati del telefono cellulare per modellare la dinamica delle presenze delle persone nella citt\`a}}
\author{Rodolfo Metulini and Maurizio Carpita}
\institute{Rodolfo Metulini \at DMS StatLab - Department of Economics and Management, University of Brescia, Contrada Santa Chiara, 50,
	25122 Brescia, \email{rodolfo.metulini@unibs.it}
	\and Maurizio Carpita \at DMS StatLab - Department of Economics and Management, University of Brescia, Contrada Santa Chiara, 50, 25122 Brescia, \email{maurizio.carpita@unibs.it}
	\and \textbf{Please cite as:} Metulini, R., Carpita, M. (2019) The HOG-FDA Approach with Mobile Phone Data to Modeling the Dynamic of People's Presences in the City. Statistical Methods for Service Quality
	Evaluation: Book of short papers of IES 2019, Rome, Italy, July 4-5, editors: Matilde Bini, Pietro Amenta, Antonello D'Ambra, Ida Camminatiello.
	}
%
%
\maketitle

\abstract{In the context of Smart City, the dynamic of the presence of people can be analysed using high-dimensional spatio-temporal mobile phone data. 
In order to find regularities and detect anomalies in the daily profiles, we propose an approach that considers the spatial structure by means of Histogram of Oriented Gradients (HOG) method and the temporal evolution using a Model-Based Clustering Functional Data Analysis (FDA).
An application to the case study of the Municipality of Brescia is provided.
Similarities among days, that follow a seasonal or a days of the week trend, exist.
The number of users in the city, depending on the season, the day of the week and the time of the day, varies from 30 to 60 thousands of people.}
\abstract{\emph{Nel contesto ``Smart City", la dinamica della presenza di persone in una certa area pu\`o essere analizzata utilizzando dati di telefonia mobile ad alta dimensionalit\`a.
Per identificare regolarit\`a e anomalie nei profili giornalieri, proponiamo un approccio che considera sia la struttura spaziale, mediante un metodo di clustering di immagini, sia l'evoluzione temporale, utilizzando un metodo di ``model-based clustering" per dati funzionali.
Applichiamo il metodo ad un caso di studio relativo al Comune di Brescia.
Troviamo somiglianze tra i giorni, che seguono una tendenza stagionale o  settimanale.
Il numero di utenti in citt\`a, a seconda della stagione, del giorno della settimana e dell'ora del giorno, varia da 30 a 60 mila persone.}}
\section{Introduction}
\label{sec:intro}
Monitoring the dynamic of the presences in the cities is crucial for the study of the well-being of an urban area in the context of ``Smart-Cities".
We use mobile phone data provided by Telecom Italia Mobile (TIM),
available thanks to a research collaboration of \href{https://sites.google.com/a/unibs.it/dms-statlab/}{DMS StatLab} with the Statistical Office of the Municipality of Brescia.
The data refer to the mobile phone activity recorded in the period September 1st 2015 to August 11th 2016 at intervals of 15 minutes in a rectangular region defined by latitude 45.516$^{\circ}$ N - 46.564$^{\circ}$ N and longitude 10.18$^{\circ}$ N - 10.245$^{\circ}$ N (Municipality of Brescia). 
Data were aggregated into 39 x 39 cells of 150 $m^2$ size each.  
For each cell and for each time interval, the corresponding record refers to the average number of mobile phones connected in that area in that time interval. 
Similar data has been used by Carpita and Simonetto \cite{carpita14big}, Zanini et al. \cite{zanini16understanding}, Metulini and Carpita \cite{metulini18on}, and Secchi et al. \cite{secchi17analysis}.

With the aim of estimating the presence of users in a specific region and over the time and to produce ``reference" daily density profiles (DDP, curves representing the number of people in a rectangular region evaluated during a day), we propose a procedure that classifies similar days using a mix of traditional (k-means) and model-based functional data clustering techniques .
When applying the procedure we find a clear separation among groups of days.
The number of users in the city of Brescia depends on the season, the day of the week and the time of the day, varying from 30 to 60 thousands of people and reaching the pick during the morning and the afternoon hours.
\section{The HOG-FDA Procedure}
\label{sec:proc}
Firstly, raw data undergoes to a strategy of dimensionality reduction. 
For each day $i$ and for each quarter $t$, we extract relevant features from the matrix of cell's values using ``Histogram of Oriented Grandients" (HOG) method \cite{dalal2005histogram}. 
Metulini and Carpita \cite{metulini18on} for details.

We then perform a traditional $k-means$ cluster analysis \cite{hartigan79algo} with the aim of grouping days that are similar in terms of the matrix of cell's values.
In detail:
\begin{enumerate}	
\item Let $i$ be a day,  $\forall \ i$ we stack (in row) into a single vector the relevant features of all the quarters, producing the vectors $f_{i}$;
\item we form the matrix $F$ = $[f_{1}, f_{2}, ..., f_{i}, ...]$ by stacking (in column) all the $f$s;
\item a $k-means$ cluster analysis is applied to group days ($F$'s columns) in terms of HOG features ($F$'s rows);
\item number of groups is chosen by looking to the decreasing of the \textbf{$\frac{within \ deviance}{total \ deviance}$} by increasing the number of groups.
\end{enumerate}		
Then, for each cluster, we perform the $model$-$based$ functional data analysis (FDA) clustering technique proposed by Bouveyron and Come \cite{bouveyron15discriminative} on DDPs to further group days. 
The aim here is to group days that are similar in terms of the dynamic of the DDP over the quarters.
In detail:
\begin{enumerate}
\item For each cluster, we remove abnormal DDPs using functional data outlier detection by likelihood ratio test (LRT), as proposed by Febrero-Bande et al. \cite{febrero08outlier};
\item we apply the method developed by Bouveyron et al. by smoothing DDP curves with a Fourier basis and by using BIC to choose for the number of groups.
\end{enumerate}
Lastly we define confidence intervals for DDPs using functional box plots \cite{sun11functional}, the analog of the traditional box plot for curves, for each final sub-cluster.
\section{An application to the city of Brescia}
\label{sec:appl}
\begin{figure}[htbp!]
\sidecaption
\includegraphics[scale=0.22]{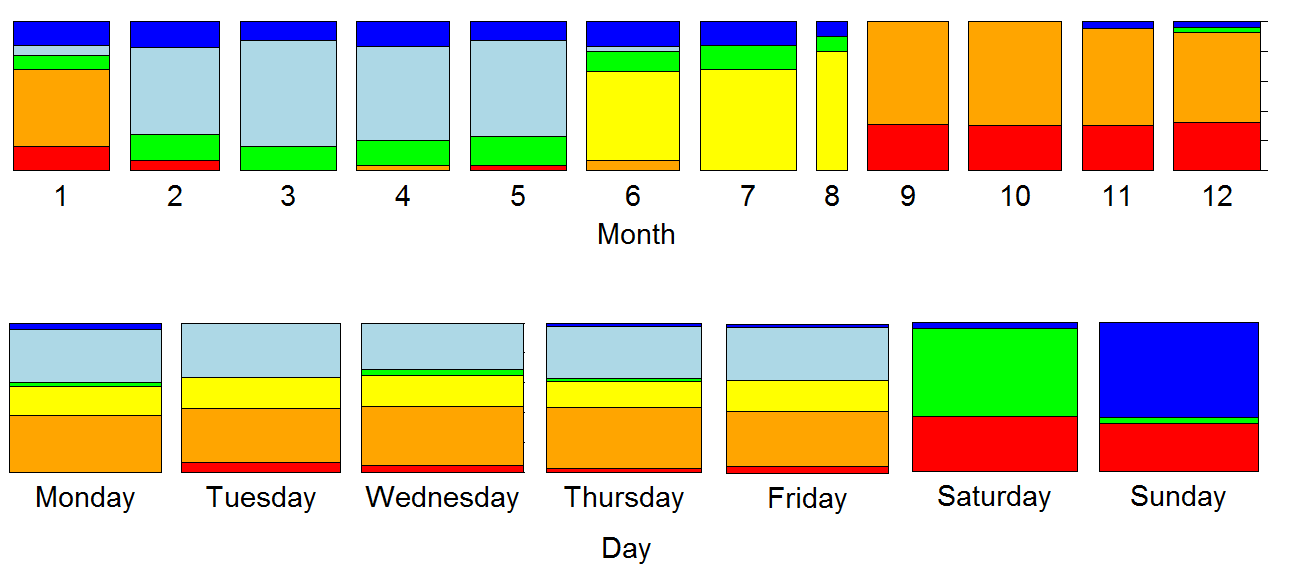}	
\caption{Distribution of the clusters by month and by week day; c1: red, c2: orange, c3: yellow, c4: green, c5: light blue, c6: blue)}
	\label{fig:spine}
\end{figure}

\begin{figure}[htbp!]
\sidecaption
\includegraphics[scale=0.21]{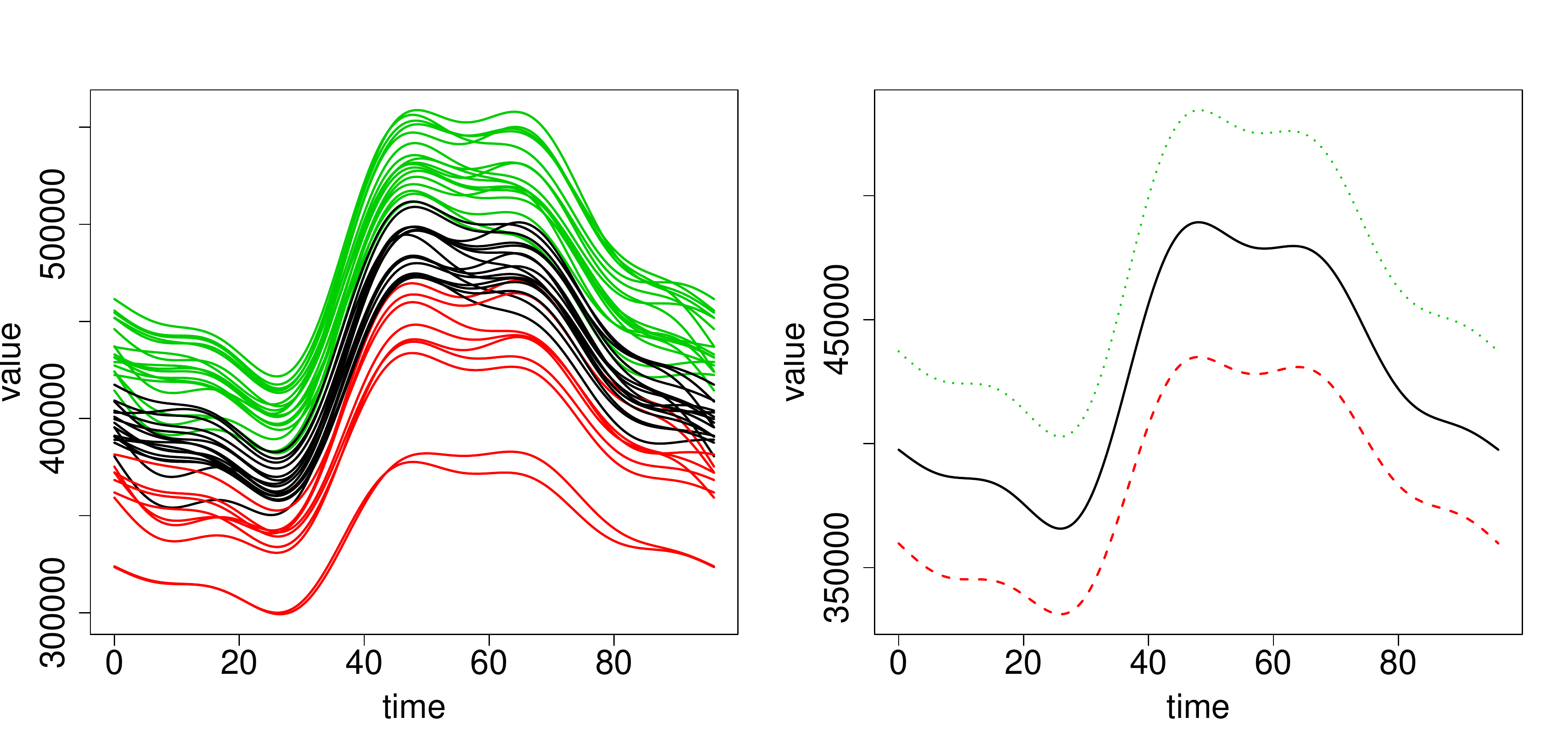}
\caption{Estimated DDP by FDA group (left) and groups' centroid (right). Week days of summer, yellow cluster (2 outliers removed with LRT). Quarters (x-axis), number of people (y-axis).}
\label{fig:fda}
\end{figure}

\begin{figure}[htbp!]
	\sidecaption
\includegraphics[scale=0.21]{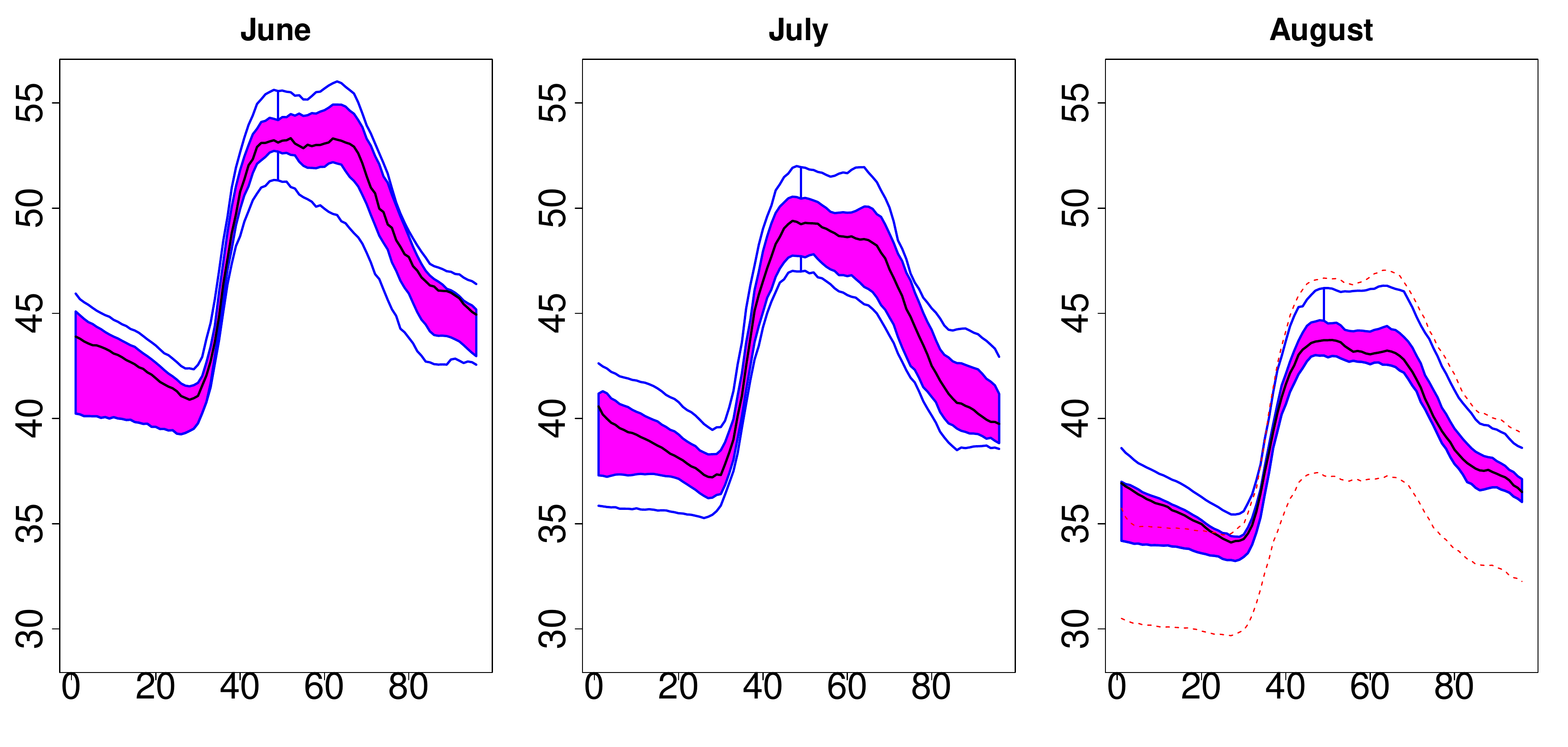}
	\caption{Functional box plot with estimated DDPs with bandwidth for (from left to right) the June, the July and the August clusters. Quarters (x-axis), thousand of people (y-axis).}
	\label{fig:fbp}
\end{figure}

After having handled missings we apply the procedure to the city of Brescia based on 330 days.
K-means cluster splits days in six well separated groups (Fig.\ref{fig:spine}). 
We report the model-based FDA clustering results for the yellow cluster with June to August and Monday to Friday days, Fig.\ref{fig:fda}. 
The method splits days in three groups (green, black and red) that correspond, respectively, to June, July and August.
The plot in Fig.\ref{fig:fbp} reports estimated DDPs with bandwidth. 
The number of people is greater in the central hours of the days and in June rather than in August.
\section{Conclusions}
\label{sec:conc}
In this work we proposed a procedure that adopts a model-based functional data clustering approach to find ``reference" daily profiles in terms of the presence of people in the city along the day time.
In future analysis we aim at setting-up a procedure to estimate the total number of users, not just those with a TIM smartphone.
\begin{acknowledgement}
	Authors are grateful with the Statistical Office of the Municipality of Brescia, with a special mention to Dr. Marco Palamenghi, Dr. Paola Chiesa and Dr. Maria Elena Comune, who kindly supported us with providing the data. 
\end{acknowledgement}
%
%
%
%

\end{document}